\documentclass{article}

 \usepackage[final]{neurips_2022}

\usepackage[utf8]{inputenc} 
\usepackage[T1]{fontenc}    
\usepackage{hyperref}       
\usepackage{url}            
\usepackage{booktabs}       
\usepackage{amsfonts}       
\usepackage{nicefrac}       
\usepackage{microtype}      
\usepackage{xcolor}         
\usepackage{graphicx}
\usepackage{natbib}
\setcitestyle{authoryear,open={(},close={)}} 
\usepackage{caption}
\usepackage{subcaption}

\title{A Study of Human-Robot Handover through Human-Human Object Transfer} 

\author{%
  Charlotte Morissette\thanks{All authors are with Samsung AI Center Montreal, 1000 Sherbrooke St W, Montreal, Qc H3A 3G4} \thanks{Gregory Dudek and Charlotte Morissette are with the school of computer science at McGill University.}\\
  \texttt{c.morissette@partner.samsung.com} \\
  \And
  Bobak H. Baghi \footnotemark[1] \\
  \texttt{ bobak.h@samsung.com} \\
  \AND
  Francois R. Hogan \footnotemark[1] \\
  \texttt{f.hogan@samsung.com} \\
  \And
  Gregory Dudek \footnotemark[1] \footnotemark[2]\\
  \texttt{ greg.dudek@samsung.com} \\
}

\begin{document}

\maketitle

\begin{abstract}
In this preliminary study, we investigate changes in handover behaviour when transferring hazardous objects with the help of a high-resolution touch sensor. Participants were asked to hand over a safe and hazardous object (a full cup and an empty cup) while instrumented with a modified STS sensor. Our data shows a clear distinction in the length of handover for the full cup vs the empty one, with the former being slower. Sensor data further suggests a change in tactile behaviour dependent on the object’s risk factor. The results of this paper motivate a deeper study of tactile factors which could characterize a risky handover, allowing for safer human-robot interactions in the future.

\end{abstract}

\section{Introduction}

In this paper, we consider the exchange of objects between humans and robots, as well as the contextual factors that modulate this behaviour. We do this by observing some of the factors that govern interhuman handover and speculate on how this can and should impact human-robot interactions. 

As robots integrate into society, ensuring safe and efficient human-robot interactions will be crucial. We envision a future where service robots assume the role of baristas and waiters who must reliably exchange objects with clientele. The transfer of an object from one person to another is a common human-human interaction that occurs in our daily lives and involves complex implicit cues and interaction dynamics. The subject of human-human handover has received some attention from researchers in recent years, where humans are found to follow a set of unspoken rules in order to successfully transfer an object. Humans seamlessly adapt their behaviour to changes in object shape, texture, size, etc., without compromising the effectiveness of the handover. The purpose of this study is to identify the key factors that change during a hazardous handover. Equipped with this understanding, our long term objective is to develop robotic behaviours that leverage this knowledge for safe human-robot object handover. 

The successful transfer of an object between agents requires a high level of reasoning and integration of low-level feedback. Humans are experts at inferring intent to effectively predict the location of handover while at the same time monitoring visual and tactile cues to understand how to retrieve and/or release the object from their partner. Inspired by this, we aim to build an understanding of the guiding principles that govern handover by studying human-human interaction. During handover, what are the key factors that inform the release time of an object? What are the types of feedback exploited to infer the release time and how to negotiate the object transfer process?

We focus our attention on the differences between how humans transfer safe versus potentially harmful objects. While past research has studied several aspects of human-human handover, such as the effects of grip force on the interaction \citep{dohring2020grip, chan2012grip, mason2005grip}, human preference for a particular handover configuration \citep{cakmak2011human}, and the role of social context in shaping action kinematics \citep{becchio2010toward}, little is known 
regarding how we exchange objects with a high failure risk. For example, when a robot passes you a scalding hot cup of coffee, a flask of liquid nitrogen, or 
a glass of expensive champagne, how should the nature of handover vary? We hypothesize that the laws that govern successful object transfer are different for scenarios that involve increased levels of danger. For example, when manipulating a  hazardous object we will more closely monitor the forces applied on the cup by the other participant to ensure that they have secured a solid grasp on the object prior to release. The objective of this research is to identify what are the key sensory factors that inform successful object transfer and how these differ for different object types.

In this study, we propose a methodology to study human-human handover using a haptic hand-held tactile device. We design a wrist-mounted tactile enabled finger that allows researchers to collect large datasets of human-human handover interactions while recording visual and tactile sensory feedback. Our hand-held device integrates the STS sensor by  \cite{hogan2021seeing}, an optical-based tactile sensor that renders high resolution images of the contact surface geometry. When an object is pressed against the STS sensor, a camera located inside the finger of the gripper captures the deformation of the sensor’s membrane and is informative of the applied contact forces and locations on the object being transferred. During handovers, the sensor records at a high frequency a series of rich tactile imprints that are informative of the cues that guide humans during object handover. In this study, we aim to determine the visual and tactile factors that best characterizes what we refer to as the manipulatory negotiation phase (i.e. the moment when both the passer and the receiver are in contact with the item). We find that this phase produces a rich tactile feedback signal, and can inform us about the release timing of an object.

This paper is organized as follows. In Section~\ref{sec:related_work}, we outline the previous research relevant to our study. In Section~\ref{sec:hand_held_device}, we introduce a novel hand-held haptic device used to collect tactile signatures during human-human handover experiments. In Section~\ref{sec:data_collection}, we summarize the experimental procedure used to collect human-human handovers while recording visual and high-resolution feedback. Finally, in Section~\ref{sec:experimental}, we relate our findings suggesting that handover behaviour is in fact affected by object risk factor.

\section{Related Work}
\label{sec:related_work}
Human-robot handover is a growing area of research, where many are studying human-human interactions for insights on how to transfer this behaviour to robots.

\subsection{Human-Human Handover}
There is a wealth of literature that studies how humans perform handover on a daily basis. Research by \citet{chan2012grip} and \citet{mason2005grip} looks into the role played by force feedback during handover. The aim of these studies is to better understand how grip force changes between the roles of the receiver and the passer as well as how changes in load forces affect grip. 

Other works have focused on the inference of grasp intentions, such as how body movements can provide important information about imminent grasp \citep{becchio2012grasping}. There is evidence that gaze cues give information about handover timing and can help improve human-robot object transfer \citep{moon2014meet}. Understanding intentions from human behaviour is increasingly important for human-robot interactions in order to reduce uncertainty in the robots perceived intent. The effects of uncertainty on the interactions of both passer and receiver are also analysed in a study by \cite{endo2012haptics}. \cite{becchio2010toward} describes how humans are social beings that spend most of their time in social settings and study how social context plays a pivotal role in shaping action planning in handover. They show that handover behaviour is shaped not only by the involved individuals but also by  the environment and context.

\subsection{Human-Robot Handover}
Numerous existing works in the human-robot handover domain draw their inspiration from the study of human-human interactions.
Recent works such as those by \cite{sanchez2020benchmark} and \cite{chao2022handoversim} focus on creating benchmarks for human-robot object transfer. The benchmark by \citet{sanchez2020benchmark} simulates the passer's behaviour and creates training and test environments for the receiver. 

Others have focused on the human side of robot transfer to ensure smooth interactions between humans and robots. \citet{cakmak2011human} studied robotic handover configurations that are preferred by humans. The question of fluency in object transfer is also addressed in a paper by \cite{cakmak2011using}, which aims to close the gap between human-robot and human-human handovers.

A natural approach to the design of human-robot handover is to study and replicate human-human handover behaviour. \citet{strabala2013toward} study human-human handover for the purpose of codification into a set of handover rules. An analysis of human collaboration was conducted to investigate how robots might be able to achieve the same level of fluency as humans in collaboration tasks \citep{huang2015adaptive}. Furthermore, collaboration or joint action is an important aspect of a safe and trustworthy robot handover, it was found that integrating knowledge of human-human joint action improves the success rate of human-robot handover \citep{grigore2013joint}.

\subsection{Tactile Handover}

An important part of handover is the object transfer phase, which begins after the receiver makes contact with the object and both participants begin negotiating the transfer. It has often been hypothesized that haptic feedback plays an important role in this phase of handover. Research by \cite{endo2012haptics} corroborates this hypothesis, finding that haptic feedback is key in timing the transfer. Other papers have looked into tactile signals and their role in handover \citep{cabibihan2013tactile}.  With a growing understanding of the importance of tactile information in human-human handover people have looked to tactile sensors to transfer this behaviour to robots. Research by \cite{yamaguchi2017implementing} work to implement manipulation strategies using a vision based tactile sensor that allows for different measures such as pose, shape and slip. Other works have looked to various tactile sensors for manipulation task, implementing diverse mechanisms to approach human dexterity \citep{kakani2021vision}.

\section{In-Hand Haptic Device}
\label{sec:hand_held_device}

\begin{figure*}[htbp]
    \centering
    \begin{subfigure}[b]{0.39\textwidth}
        \includegraphics[width=\textwidth, height=0.29\textheight]{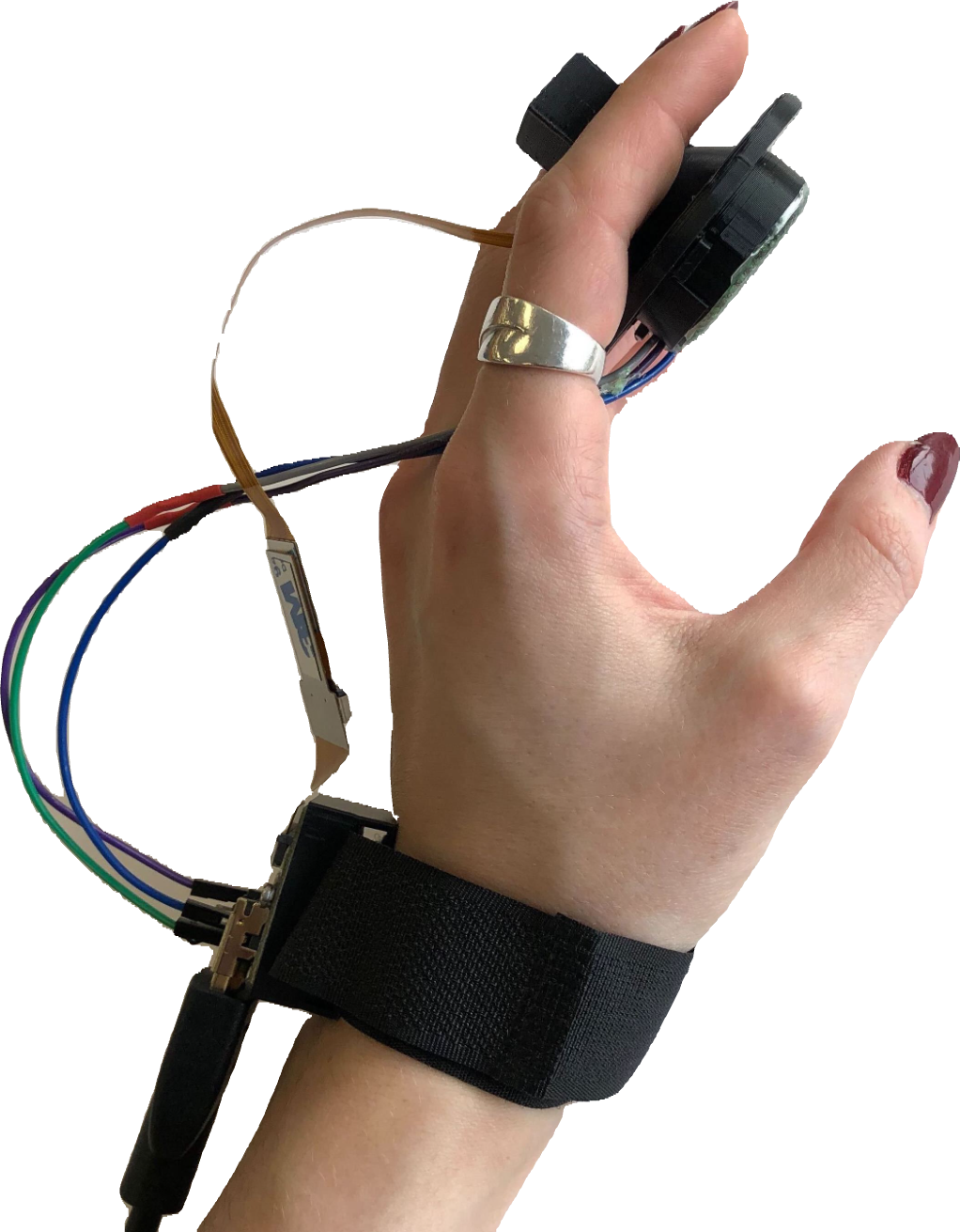}
        \caption{}
    \end{subfigure}
    \hfill
    \begin{subfigure}[b]{0.39\textwidth}
        \includegraphics[width=\textwidth, height=0.29\textheight]{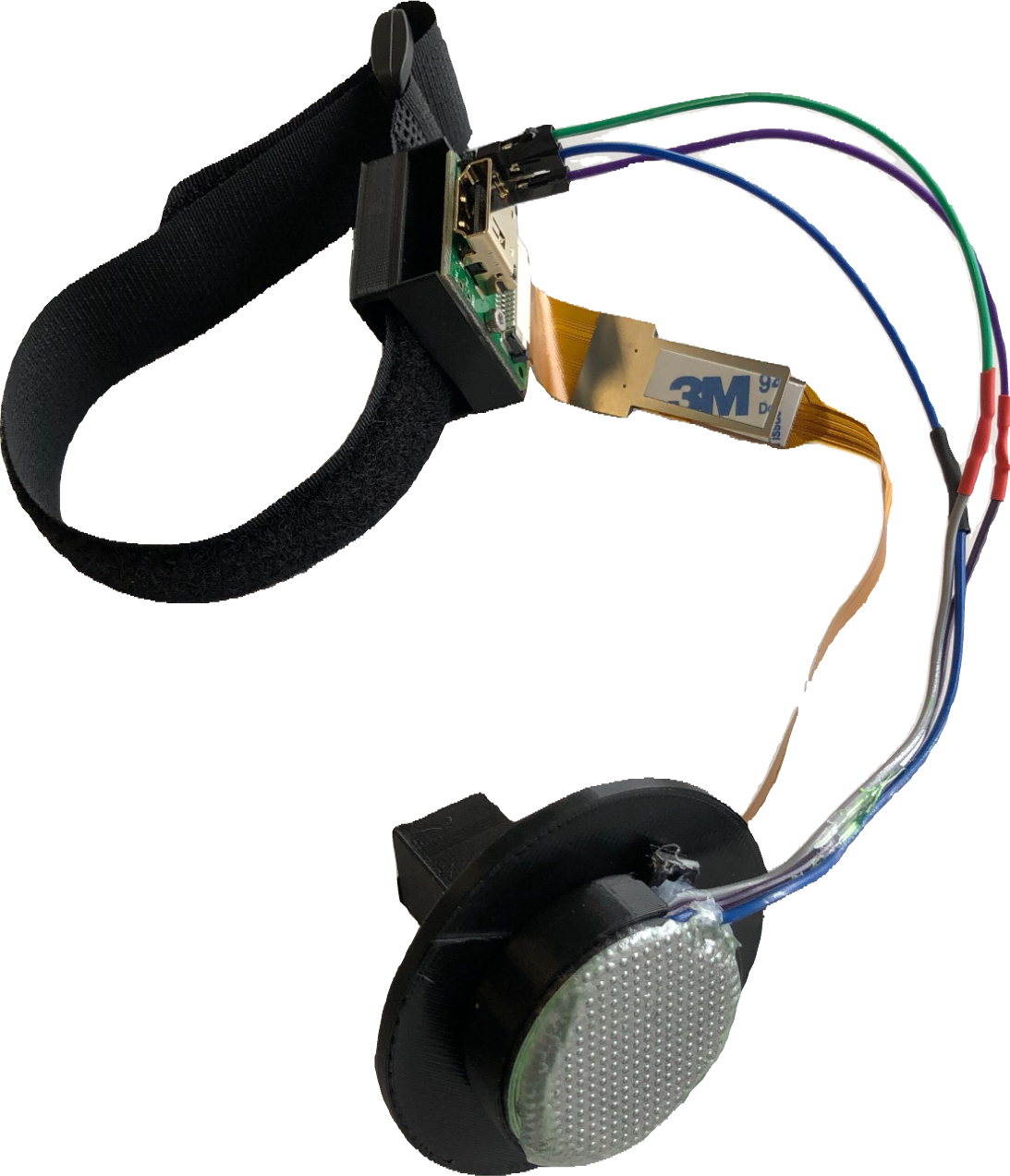}
        \caption{}
    \end{subfigure}
    \caption{In-hand haptic device, adapted version of See-Through-your-Skin (STS) visuotactile sensor.}
    \label{fig:hardware_setup}
\end{figure*}

In this work, we fabricate and use a high-resolution tactile sensor to characterize the nature of contact during human-human handovers. We design and adapt a version of the See-Through-your-Skin (STS) visuotactile sensor \citep{hogan2021seeing} to be held by a human hand and used for natural object handover. 

The operating principle of this sensor is to capture the deformations of a gel membrane using a camera. To amplify the visibility of these deformations, the gel membrane is coated in a thin, partially reflective layer and internally lit using an array of LEDs. Furthermore, the membrane is adorned with a pattern of regularly spaced circular dots, which deform along with the membrane. These markers can be easily tracked by the camera independently of the background image that the camera sees. The camera, LEDs, and membrane form a unit that is held together with a rigid shell. The necessary computation and image processing are performed by a Raspberry PI that is connected to the sensor via HDMI cable. The assembly is shown in \autoref{fig:hardware_setup}.

The thin coat of silver paint on the sensor’s membrane allows for multi-modal visuotactile sensing. Similar to a one-way mirror, changing the lighting changes the membrane, making it either see-through, allowing for visual feedback, or reflective, allowing for tactile imprints to be recorded. This visual-tactile property can be leveraged to record different physical attributes such as slip, texture, proximity, shape and force. \citep{hogan2021seeing} 

Since we wish to use the sensor for human handover, the paramount guiding principle in this design is to minimize the impact of the handheld sensor on natural human-human handover behaviour. We accomplish this by adapting the shape of the rigid shell so that it can be held naturally between the index and middle finger on one hand, allowing the user to still experience most of the haptic feedback involved in an object handover. The participants wore a wristband which constrained the electronics in place and served as a wire guide so as not to make them interfere with the handover process.

\section{Data Collection}
\label{sec:data_collection}
We report observations from two datasets, both of human-human handover, the first where one participant wears the visuotactile sensor and a control study where neither participants had sensors. The set up for both these experiments is identical. Two participants sat across from each other seperated by a table. The table is set up in the following way; on its surface there were three square areas marked with masking tape, the first on the ledge at one side of the table in front of participant A. The second is at the opposite side of the table in front of participant B. The third and final square was drawn equidistant between the first two. In order to record the handovers, a USB camera was placed on the side of the table vis-à-vis the middle square, filming the interactions from above.

In both experiments, two objects were used for handover. One was an empty 8oz paper cup, and the other was the same cup filled to the brim with water. The latter object constitutes a hazardous handover where more caution is required to avoid any spillage. The choice of handing over two identical cups with drastically different contents was to analyze how handover behaviour changes when going from a safe/easy handover (empty cup) to a riskier one (full cup).

At the start of the experiment, an object is placed in front of participant A in the square drawn in from of them, a signal is then given for them to pick up the object but not to begin handover. Once the object is picked up a signal is given for participant A to begin handover, they are now permitted to move the object out of the square in front of them to the middle square where handover can occur. Participant A, the passer, and participant B, the receiver, meet in the center square for a phase dubbed the negotiation phase where both participants are in contact with the object. Once the passer has released the object the receiver puts down the object in the square facing them marking the end of the handover. The process then begins again with participant B now in the passer role and participant A as the receiver. This procedure is identical for both of the datasets we collect.

We annotate the collected datasets manually using the accompanying top-down camera view of the experiment to mark the following moments: When either participant starts the handover process (start of horizontal movement following the signal), when the two participants are both touching the object (negotiation phase), the moment when one subject releases the object, and finally the moment when the handover is complete and the object is brought to the staging area of the participant.

\section{Experimental Results}
\label{sec:experimental}

\subsection{Human-Human sensor free handover}
\begin{figure}[htbp]
    \centering
    \begin{subfigure}[b]{0.49\textwidth}
        \includegraphics[width=\textwidth]{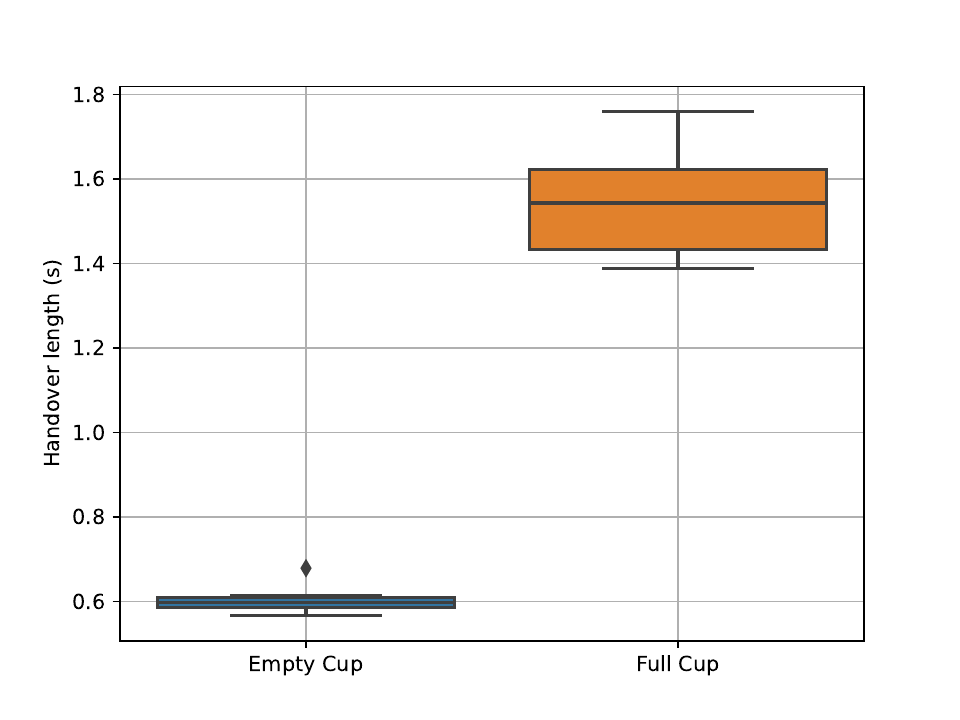}
        \caption{Duration of handover.}
    \end{subfigure}
    \hfill
    \begin{subfigure}[b]{0.49\textwidth}
        \includegraphics[width=\textwidth]{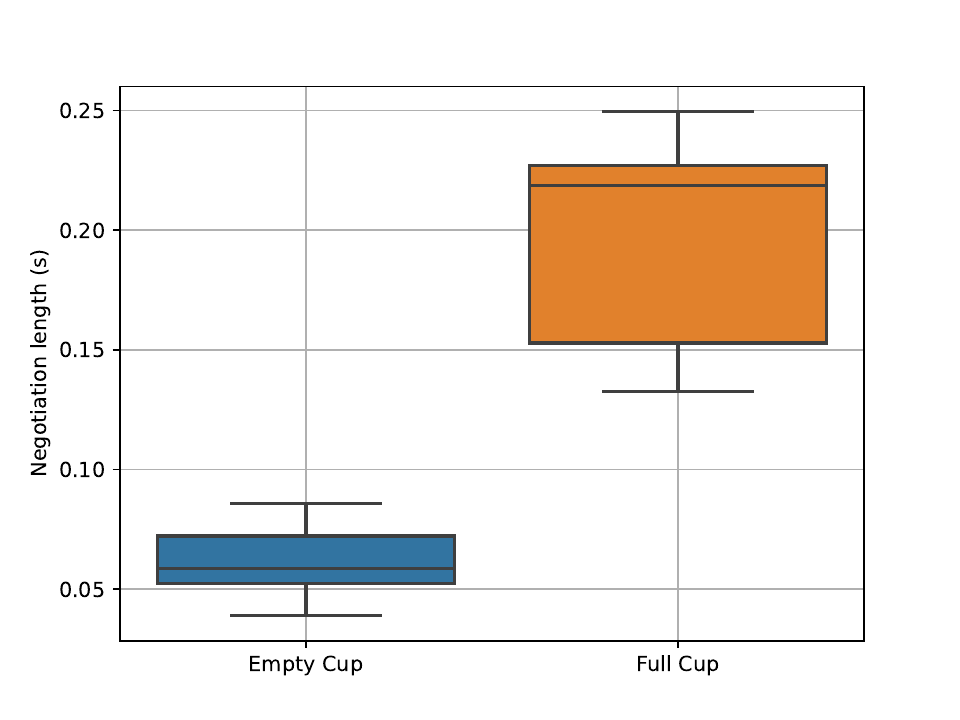}
        \caption{Duration of negotiation phase.}
    \end{subfigure}
    \caption{Behavioral differences in the handover of hazardous (full cup) and safe (empty cup) in control experiment (without STS instrumentation). The duration of the total handover (a) and the duration of the negotiation phase (b) are longer in the hazardous (full cup) case.}
    \label{fig:behaviour_differences}
\end{figure}

In this experiment, two participants engage in handover without the use of a visuotactile sensor. We observed that the participants showed increased caution in the handover of the hazardous object, namely the full cup. As can be seen in \autoref{fig:behaviour_differences}, the total handover duration as well as the negotiation phase duration is longer in the full cup handover. We attribute this to the increased caution required to perform the full cup handover, in particular, the increased constraints; the full cup must be kept upright at all costs which significantly limits mobility. Qualitatively, the negotiation phase of the full cup was performed much more carefully to constrain the rotation of the cup.

\subsection{Handover Using Visuotactile Sensor}

\begin{figure}[htbp]
    \centering
    \begin{subfigure}[b]{0.45\textwidth}
         \centering
         \includegraphics[width=\textwidth]{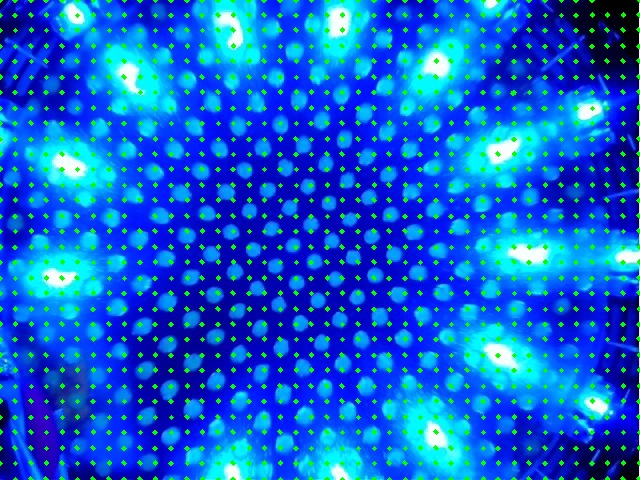}
         \caption{Tactile measurement from STS sensor when no deformation is present.}
         \label{fig:noflow}
     \end{subfigure}
     \hfill
     \begin{subfigure}[b]{0.45\textwidth}
         \centering
         \includegraphics[width=\textwidth]{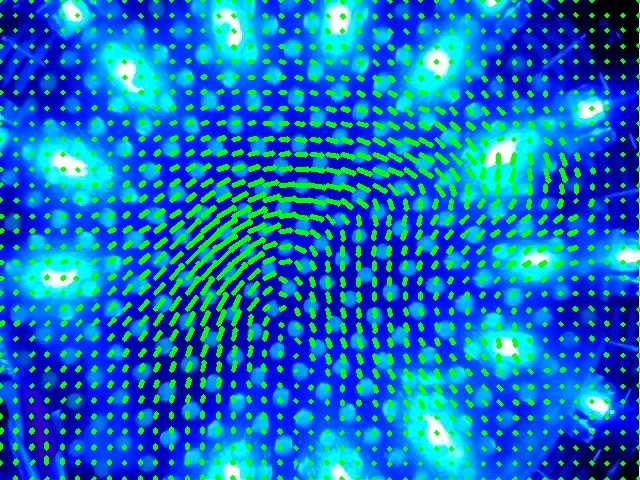}
         \caption{Tactile measurement from STS sensor with deformation shown by the green optical flow vectors.}
         \label{fig:flow}
     \end{subfigure}
    \caption{STS generated tactile measurement. The overlaid green vectors show a subsampling of the dense optical flow computed at each pixel. In the left figure (a), there is almost no deformation. In (b), there is some rotational deformation resulting in the optical flow vectors as shown.}
    \label{fig:sts_flow}
\end{figure}

\begin{figure}[htbp]
     \centering
     \begin{subfigure}[b]{0.3\textwidth}
         \centering
         \includegraphics[width=\textwidth]{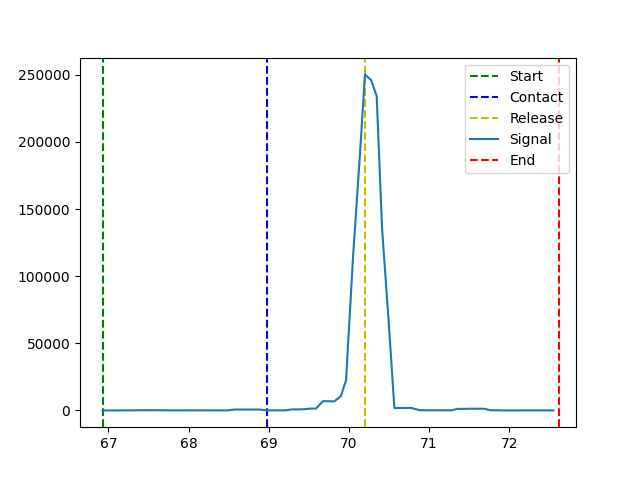}
         \caption{}

     \end{subfigure}
     \hfill
     \begin{subfigure}[b]{0.3\textwidth}
         \centering
         \includegraphics[width=\textwidth]{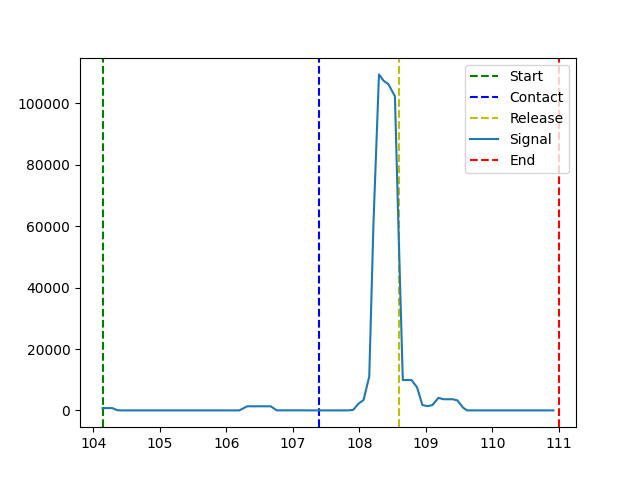}
         \caption{}

     \end{subfigure}
     \hfill
     \begin{subfigure}[b]{0.3\textwidth}
         \centering
         \includegraphics[width=\textwidth]{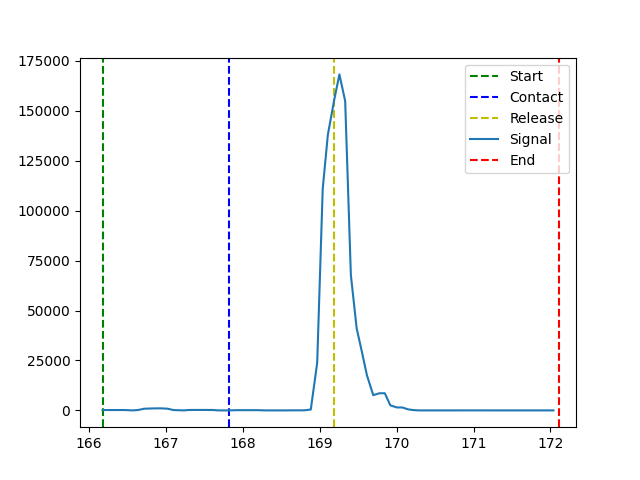}
         \caption{}

     \end{subfigure}
        \caption{Full cup handover curves. The x-axis is time in seconds, while the y-axis is the sum of all optical flow vector magnitudes for that frame in the STS sensor reading. Figures (a-c) each signify an individual handover. A sharp increase in the optical flow magnitudes can be observed at the release moment.}
        \label{fig:full-cup-curves}
\end{figure}

\begin{figure}[htbp]
     \centering
     \begin{subfigure}[b]{0.3\textwidth}
         \centering
         \includegraphics[width=\textwidth]{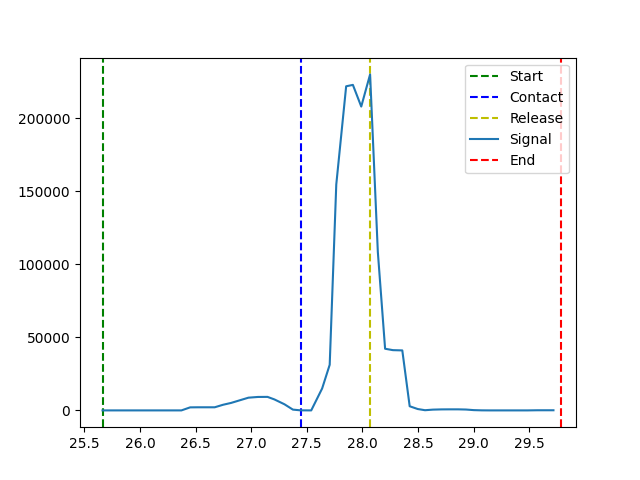}
         \caption{}
         \label{fig:y equals x}
     \end{subfigure}
     \hfill
     \begin{subfigure}[b]{0.3\textwidth}
         \centering
         \includegraphics[width=\textwidth]{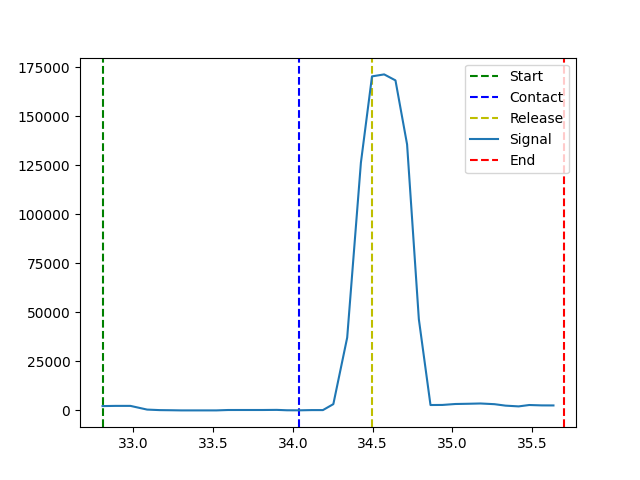}
         \caption{}
         \label{fig:three sin x}
     \end{subfigure}
     \hfill
     \begin{subfigure}[b]{0.3\textwidth}
         \centering
         \includegraphics[width=\textwidth]{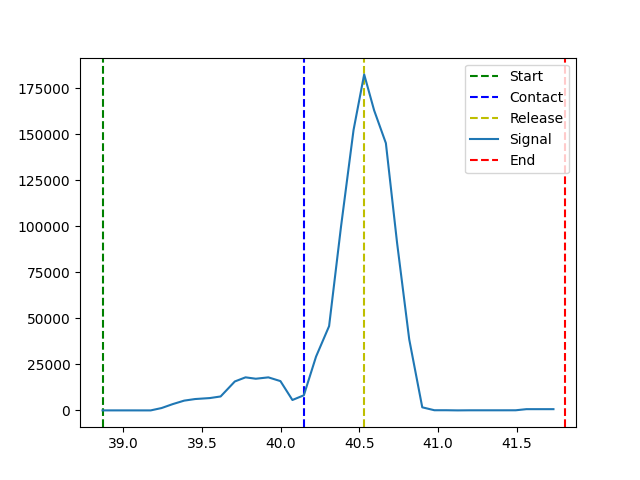}
         \caption{}
         \label{fig:five over x}
     \end{subfigure}
        \caption{Empty cup handover curves. The x-axis is time in seconds, while the y-axis is the sum of all optical flow vector magnitudes for that frame in the STS sensor reading. Figures (a-c) each signify an individual handover. An increase in the optical flow magnitudes can be observed at the release moment.}
        \label{fig:empty-cup-curves}
\end{figure}

From the human-human handover experiments without sensor we noticed a clear increase in handover length when manipulating a riskier object, such as the full cup, these results motivated us to look more closely at the transfer and negotiation phase of the handover to determine which tactile factors might be changing from simple to risky object transfer. In these experiments, one participant wore the in-hand haptic sensor to record tactile data. While the STS sensor is capable of recording rich tactile information, we focused solely on quantifying the magnitude of membrane deformation as a loose proxy for the severity of the negotiation phase. Specifically, we compute dense optical flow vectors of the STS image readings(see \autoref{fig:flow}), and compute their per-frame cumulative magnitude as a measure of total surface deformation, which can be seen in the \autoref{fig:full-cup-curves} and \autoref{fig:empty-cup-curves}. The green line indicates the start of handover, the blue the moment the receiver makes contact with the object, the yellow the release phase and red the end of handover. Each figure has three different instances of handover.

In both the safe and hazardous cases, a spike near the moment of release and after the start of the negotiation phase can be observed. This is expected, as it is before release that the participants verify the grasp stability of their counterpart. We observe, as in the sensor-free experiment discussed above, that the hazardous object handover has longer duration, therefore reinforcing the notion that hazardous handover requires a longer time window. However, of particular note, is the qualitative properties of the flow magnitude spikes. As can be seen in the  \autoref{fig:full-cup-curves} and \autoref{fig:empty-cup-curves},  the magnitude spike for the full cup is much more narrow and sudden than the empty cup. This indicates less deformation and therefore a more careful handover for the hazardous case. It also confirms the existence of a tactile component to handover detectable by the STS which could not be explored solely in the visual domain with an experiment design as discussed above. 

Overall, this result motivates further and more fine-grained investigation of the tactile signatures that inform handover through the use of visuotactile sensors such as the STS.

\section{Summary and Future Work}
This work constitutes a preliminary study of the means by which a visuotactile sensor can be useful in the handover process. In particular, we focus our study on the distinctions between handover of high-risk vs. safe objects and the constraints that govern them. We show preliminary qualitative and quantitative evidence of differences in these cases. This includes a noticeable increase in handover duration in riskier object transfers as well as more careful handovers for hazardous ones as seen in the sensor data. 

In future work, we aim to expand on this direction. An interesting future direction is a more salient identification of features that predict the moment when a handover might commence or end. Further study can enable the transfer of handover behaviour to a real robot, perhaps through an imitation strategy.

\bibliography{main.bib}

\begin{thebibliography}{18}
\providecommand{\natexlab}[1]{#1}
\providecommand{\url}[1]{\texttt{#1}}
\expandafter\ifx\csname urlstyle\endcsname\relax
  \providecommand{\doi}[1]{doi: #1}\else
  \providecommand{\doi}{doi: \begingroup \urlstyle{rm}\Url}\fi

\bibitem[Becchio et~al.(2010)Becchio, Sartori, and
  Castiello]{becchio2010toward}
C.~Becchio, L.~Sartori, and U.~Castiello.
\newblock Toward you: The social side of actions.
\newblock \emph{Current Directions in Psychological Science}, 19\penalty0
  (3):\penalty0 183--188, 2010.

\bibitem[Becchio et~al.(2012)Becchio, Manera, Sartori, Cavallo, and
  Castiello]{becchio2012grasping}
C.~Becchio, V.~Manera, L.~Sartori, A.~Cavallo, and U.~Castiello.
\newblock Grasping intentions: from thought experiments to empirical evidence.
\newblock \emph{Frontiers in human neuroscience}, 6:\penalty0 117, 2012.

\bibitem[Cabibihan et~al.(2013)Cabibihan, Wu, and
  Ramalingam]{cabibihan2013tactile}
J.-J. Cabibihan, K.~W. Wu, and A.~Ramalingam.
\newblock Tactile sensing in an object passing task.
\newblock In \emph{2013 IEEE Conference on Cybernetics and Intelligent Systems
  (CIS)}, pages 96--99. IEEE, 2013.

\bibitem[Cakmak et~al.(2011{\natexlab{a}})Cakmak, Srinivasa, Lee, Forlizzi, and
  Kiesler]{cakmak2011human}
M.~Cakmak, S.~S. Srinivasa, M.~K. Lee, J.~Forlizzi, and S.~Kiesler.
\newblock Human preferences for robot-human hand-over configurations.
\newblock In \emph{2011 IEEE/RSJ International Conference on Intelligent Robots
  and Systems}, pages 1986--1993. IEEE, 2011{\natexlab{a}}.

\bibitem[Cakmak et~al.(2011{\natexlab{b}})Cakmak, Srinivasa, Lee, Kiesler, and
  Forlizzi]{cakmak2011using}
M.~Cakmak, S.~S. Srinivasa, M.~K. Lee, S.~Kiesler, and J.~Forlizzi.
\newblock Using spatial and temporal contrast for fluent robot-human
  hand-overs.
\newblock In \emph{Proceedings of the 6th international conference on
  Human-robot interaction}, pages 489--496, 2011{\natexlab{b}}.

\bibitem[Chan et~al.(2012)Chan, Parker, Van~der Loos, and Croft]{chan2012grip}
W.~P. Chan, C.~A. Parker, H.~M. Van~der Loos, and E.~A. Croft.
\newblock Grip forces and load forces in handovers: implications for designing
  human-robot handover controllers.
\newblock In \emph{Proceedings of the seventh annual ACM/IEEE international
  conference on Human-Robot Interaction}, pages 9--16, 2012.

\bibitem[Chao et~al.(2022)Chao, Paxton, Xiang, Yang, Sundaralingam, Chen,
  Murali, Cakmak, and Fox]{chao2022handoversim}
Y.-W. Chao, C.~Paxton, Y.~Xiang, W.~Yang, B.~Sundaralingam, T.~Chen, A.~Murali,
  M.~Cakmak, and D.~Fox.
\newblock Handoversim: A simulation framework and benchmark for human-to-robot
  object handovers.
\newblock \emph{arXiv preprint arXiv:2205.09747}, 2022.

\bibitem[D{\"o}hring et~al.(2020)D{\"o}hring, M{\"u}ller, and
  Joch]{dohring2020grip}
F.~R. D{\"o}hring, H.~M{\"u}ller, and M.~Joch.
\newblock Grip-force modulation in human-to-human object handovers: effects of
  sensory and kinematic manipulations.
\newblock \emph{Scientific Reports}, 10\penalty0 (1):\penalty0 1--10, 2020.

\bibitem[Endo et~al.(2012)Endo, Pegman, Burgin, Toumi, and
  Wing]{endo2012haptics}
S.~Endo, G.~Pegman, M.~Burgin, T.~Toumi, and A.~M. Wing.
\newblock Haptics in between-person object transfer.
\newblock In \emph{International Conference on Human Haptic Sensing and Touch
  Enabled Computer Applications}, pages 103--111. Springer, 2012.

\bibitem[Grigore et~al.(2013)Grigore, Eder, Pipe, Melhuish, and
  Leonards]{grigore2013joint}
E.~C. Grigore, K.~Eder, A.~G. Pipe, C.~Melhuish, and U.~Leonards.
\newblock Joint action understanding improves robot-to-human object handover.
\newblock In \emph{2013 IEEE/RSJ International Conference on Intelligent Robots
  and Systems}, pages 4622--4629. IEEE, 2013.

\bibitem[Hogan et~al.(2021)Hogan, Jenkin, Rezaei-Shoshtari, Girdhar, Meger, and
  Dudek]{hogan2021seeing}
F.~R. Hogan, M.~Jenkin, S.~Rezaei-Shoshtari, Y.~Girdhar, D.~Meger, and
  G.~Dudek.
\newblock Seeing through your skin: Recognizing objects with a novel
  visuotactile sensor.
\newblock In \emph{Proceedings of the IEEE/CVF Winter Conference on
  Applications of Computer Vision}, pages 1218--1227, 2021.

\bibitem[Huang et~al.(2015)Huang, Cakmak, and Mutlu]{huang2015adaptive}
C.-M. Huang, M.~Cakmak, and B.~Mutlu.
\newblock Adaptive coordination strategies for human-robot handovers.
\newblock In \emph{Robotics: science and systems}, volume~11, pages 1--10.
  Rome, Italy, 2015.

\bibitem[Kakani et~al.(2021)Kakani, Cui, Ma, and Kim]{kakani2021vision}
V.~Kakani, X.~Cui, M.~Ma, and H.~Kim.
\newblock Vision-based tactile sensor mechanism for the estimation of contact
  position and force distribution using deep learning.
\newblock \emph{Sensors}, 21\penalty0 (5):\penalty0 1920, 2021.

\bibitem[Mason and MacKenzie(2005)]{mason2005grip}
A.~H. Mason and C.~L. MacKenzie.
\newblock Grip forces when passing an object to a partner.
\newblock \emph{Experimental brain research}, 163\penalty0 (2):\penalty0
  173--187, 2005.

\bibitem[Moon et~al.(2014)Moon, Troniak, Gleeson, Pan, Zheng, Blumer, MacLean,
  and Croft]{moon2014meet}
A.~Moon, D.~M. Troniak, B.~Gleeson, M.~K. Pan, M.~Zheng, B.~A. Blumer,
  K.~MacLean, and E.~A. Croft.
\newblock Meet me where i'm gazing: how shared attention gaze affects
  human-robot handover timing.
\newblock In \emph{Proceedings of the 2014 ACM/IEEE international conference on
  Human-robot interaction}, pages 334--341, 2014.

\bibitem[Sanchez-Matilla et~al.(2020)Sanchez-Matilla, Chatzilygeroudis, Modas,
  Duarte, Xompero, Frossard, Billard, and Cavallaro]{sanchez2020benchmark}
R.~Sanchez-Matilla, K.~Chatzilygeroudis, A.~Modas, N.~F. Duarte, A.~Xompero,
  P.~Frossard, A.~Billard, and A.~Cavallaro.
\newblock Benchmark for human-to-robot handovers of unseen containers with
  unknown filling.
\newblock \emph{IEEE Robotics and Automation Letters}, 5\penalty0 (2):\penalty0
  1642--1649, 2020.

\bibitem[Strabala et~al.(2013)Strabala, Lee, Dragan, Forlizzi, Srinivasa,
  Cakmak, and Micelli]{strabala2013toward}
K.~Strabala, M.~K. Lee, A.~Dragan, J.~Forlizzi, S.~S. Srinivasa, M.~Cakmak, and
  V.~Micelli.
\newblock Toward seamless human-robot handovers.
\newblock \emph{Journal of Human-Robot Interaction}, 2\penalty0 (1):\penalty0
  112--132, 2013.

\bibitem[Yamaguchi and Atkeson(2017)]{yamaguchi2017implementing}
A.~Yamaguchi and C.~G. Atkeson.
\newblock Implementing tactile behaviors using fingervision.
\newblock In \emph{2017 IEEE-RAS 17th International Conference on Humanoid
  Robotics (Humanoids)}, pages 241--248. IEEE, 2017.

\end{thebibliography}
\end{document}